# Age Sensitive Hippocampal Functional Connectivity: New Insights from 3D CNNs and Saliency Mapping

Yifei Sun[1,2], Marshall A. Dalton[2,3], Robert D. Sanders[2,4], Yixuan Yuan[5], Xiang Li[6], Sharon L. Naismith[2,3], Fernando Calamante[1,2,7], Jinglei Lv[1,2,4].

1. School of Biomedical Engineering, The University of Sydney, Sydney 2050, Australia.
2. Brain and Mind Centre, The University of Sydney, Sydney 2050, Australia.
3. School of Psychology, The University of Sydney, Sydney 2050, Australia.
4. Central Clinical School, The University of Sydney, Sydney 2050, Australia.
5. Department of Electronic Engineering, The Chinese University of Hong Kong, Hong Kong, China.
6. Department of Radiology, Massachusetts General Hospital, Harvard Medical School, Boston, MA, USA.
7. Sydney Imaging, The University of Sydney, Sydney 2050, Australia.

Corresponding Author: Jinglei Lv (jinglei.lv@sydney.edu.au)

**Abstract**

Grey matter loss in the hippocampus is a hallmark of neurobiological aging, yet understanding the corresponding changes in its functional connectivity remains limited. Seed-based functional connectivity (FC) analysis enables voxel-wise mapping of the hippocampus's synchronous activity with cortical regions, offering a window into functional reorganization during aging. In this study, we develop an interpretable deep learning framework to predict brain age from hippocampal FC using a three-dimensional convolutional neural network (3D CNN) combined with LayerCAM saliency mapping. This approach maps key hippocampal-cortical connections—particularly with the precuneus, cuneus, posterior cingulate cortex, parahippocampal cortex, left superior parietal lobule, and right superior temporal sulcus—that are highly sensitive to age. Critically, disaggregating anterior and posterior hippocampal FC reveals distinct mapping aligned with their known functional specializations. These findings provide new insights into the functional mechanisms of hippocampal aging and demonstrate the power of explainable deep learning to uncover biologically meaningful patterns in neuroimaging data.

# 1. Introduction

The hippocampus plays a crucial role in various cognitive functions, including memory and spatial navigation [1, 2], yet it undergoes significant structural and functional changes with aging. One of the significant markers of brain aging is hippocampal volume reduction, which has been widely reported in neuroimaging studies [3-5]. Although these structural changes underscore that the hippocampus is sensitive to the aging progress, how it's functional connectivity (FC) with the rest of the brain changes in the context of healthy aging remains unclear. Particularly, the functional interactions of the hippocampus with other brain regions are crucial for understanding age-related cognitive decline, yet underexplored. While structural magnetic resonance imaging (MRI) studies have documented hippocampal atrophy [6, 7], resting-state functional MRI (rs-fMRI) provides a complementary perspective by capturing hippocampal FC with other brain regions [8, 9]. Seed-based FC analysis allows for a three-dimensional (3D) mapping of hippocampal synchronization patterns across the whole brain, offering high-resolution insights into its functional reorganization with age. Given the increasing global aging population, understanding these changes is essential for maintaining cognitive health and developing early interventions for age-related cognitive decline.

Machine learning, particularly deep learning, has become a powerful tool in neuroimaging-based age prediction. Early studies leveraged structural MRI with three-dimensional convolutional neural networks (3D CNNs) to estimate brain age, achieving mean absolute errors (MAE), i.e. the difference between predicted brain age and chronological age, of approximately 4 years [10]. Further refinements of CNNs for T1-weighted images improved age prediction MAE to roughly 3 years [11, 12] and to 2.14 years with the introduction of a simple fully convolutional neural network (SFCN) [13]. These relevant works show great potential for 3D CNN to learn and predict patterns of structural brain changes associated with healthy aging.

In parallel, researchers have explored FC calculated from fMRI for age prediction. Most of the explorations adopted region scale FC matrices as input, rather than voxel-wise input. One study on brain development (age range: 8–22 years) using FC matrices achieved high accuracy (MAE < 2.5 years) [14]. However, prediction accuracy reduces in aging population, likely due to larger individual variability in aging trajectories. For example, one study analysing FC matrices from 264 brain regions in adults aged 18-88 reported an MAE above 8 years [15], while another study using FC matrices from 116 regions in adults aged 51–97 achieved an MAE of 5.92 years [16]. Although these studies achieved relatively promising performance, the conventional region-based FC matrices have intrinsic limitations in their spatial resolution for capturing the fine-grained connectivity details that may be needed for accurate age estimation and explicit understanding of brain functional changes during normal aging.

This study aimed to characterise FC changes of the hippocampus associated with healthy aging through a deep learning approach. Instead of relying on coarse region-based FC matrices, we employed seed-based FC analysis focussed specifically on the hippocampus to generate whole-brain voxel-wise FC maps, providing a more detailed understanding of hippocampal connectivity patterns in the healthy aging brain. We trained a 3D CNN model (Fig.1a) to predict chronological age using hippocampal-cortical FC. We selected 3D CNN not only because of its predictive power with structural brain MRI, but also, recent technological development of class activation mapping (CAM) opens a door of interpretability of the CNN models. Particularly, layer-wise CAM (LayerCAM) [17] provides high resolution interpretability. LayerCAM was originally developed in 2D CNN for object recognition in natural images. In our study, we extended LayerCAM to 3D and adapted it for the prediction task. It helps generate a voxel-wise saliency map (Fig.1c) that highlight brain regions most influential in age estimation, enhancing transparency and interpretability.

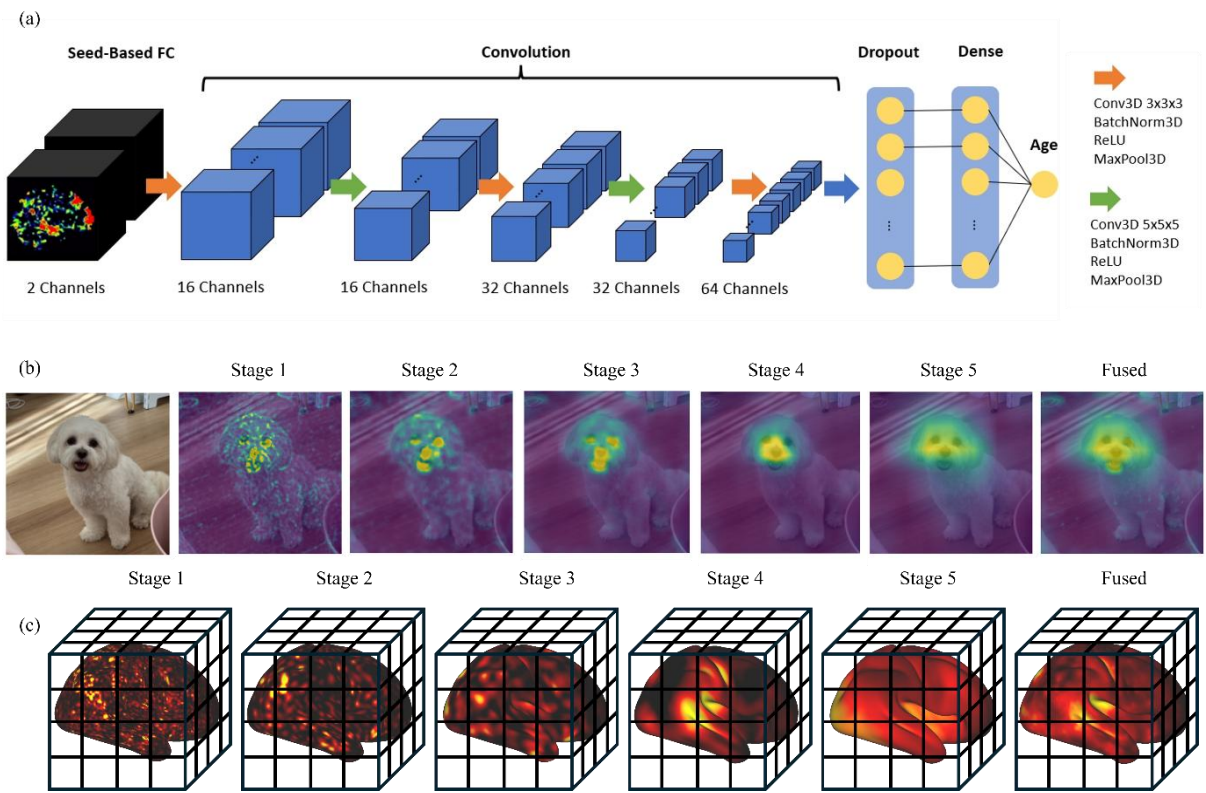

Fig. 1 (a) 3D CNN architecture used for age prediction using seed-based FC. The model stacks five convolution blocks, each containing one convolutional layer (Conv3D), one batch normalization layer (BatchNorm3D), a ReLU activation, and a max-pooling layer (MaxPool3D). Two types of convolution blocks with different kernel size are used in the convolutional layer. One of them uses the kernel of size 3x3x3 and the other one uses the 5x5x5 kernel. After convolutional blocks, a dropout layer followed by one fully connected layer is used to output the predicted age. (b) LayerCAM on natural image for object localization. This figure was recreated using the method and code from [17]. (c) LayerCAM adapted to 3D for age prediction task. The saliency map of stage 1 corresponds to the saliency map generated based on the first convolutional layer. The fused saliency map is the final heatmap after combining the maps from all 5 stages.

To further characterise age-related hippocampal FC changes, we divided the hippocampus into anterior and posterior subregions, recognizing their different connectivity patterns [18]. As reported, the anterior hippocampus is primarily associated with emotional regulation and episodic memory, while the posterior hippocampus plays a key role in spatial navigation [19-22]. By separately employing anterior and posterior hippocampal FC for age prediction, we pushed the resolution of the method and aim to uncover subregion-specific aging effects that may not be evident when treating the hippocampus as a single unit. In summary, our approach provides a novel framework for identifying key cortico-hippocampal connections associated with aging and new insights into functional brain aging mechanisms.

## 2. Results

**2.1. 3D CNN Age Prediction Performance**

We employed the human connectome project young adults (HCP-Y) [23] and aging (HCP-A) [24] dataset to cover a wide age range from 23 to 90 and guarantee a large sample number to ensure the trained model is robust. Seed-based hippocampal FC was calculated by using the left and right hippocampus as separate seeds, generating two 3D volumes to feed into the 3D CNN network. For illustration, the HCP-A group averaged hippocampal FC is represented as a 3D volume and mapped to the brain surface as shown in Fig.2b. High FC connectivity is found for several cortical areas including parahippocampal cortex (PHC), posterior cingulate cortex (PCC), medial prefrontal cortex (PFC), inferior parietal lobule (IPL), and motor cortex. Most of these regions with high FC are the main components of DMN [25], showing the strong connections between hippocampus and DMN.

Our 3D CNN model is designed as the stacking of five 3D convolutional blocks each comprising a 3D convolutional layer, a batch normalization layer, the rectified linear unit (ReLU) activation, and a maxpooling layer as shown in Fig.1a. Inspired by Inception models [26], we applied the kernel with different sizes of 3x3x3 and 5x5x5 alternatively in convolutional layers, in order to capture both fine-grained and broader global patterns from the 3D FC data. Unlike CNN models designed for classification, e.g. SFCN [13], with the SoftMax activation to generate the probability for different classes or age intervals, our CNN perform as a prediction task, therefore we replaced the SoftMax with the linear activation layer.

We designed a two-stage training and testing strategy to train a robust model. The first stage is pretraining, during which we trained these 3D CNN models on the HCP-Y dataset (age range: 22-37 years), aiming to learn the data representation, obtain a stable initial weight configuration and reduce the overfitting risk. Among 1018 subjects with all four rs-fMRI scans available, 814 subjects were used for training, and the other 204 subjects were used as a test dataset. During training and testing, we used the mean absolute error (MAE) to evaluate the model. The model predicted the age with an MAE of 2.99 years, on the HCP-Y test dataset.

Considering differences in the imaging protocol and the expected higher variability in the aging population, the second stage is to finetune the model on the HCP-A dataset with 708 subjects for domain adaptation. We mainly focused on the subjects aged 36-90, excluding 12 subjects aged above 90, because their detailed age information was not provided for privacy reason. The pretrained model with HCP-Y dataset was further trained and validated through a fivefold cross-validation with MAE as the loss function and evaluation metrics. The predictive model using the whole hippocampal cortical FC achieved the validation MAE of 7.30 years and the spearman's correlation of 0.79 between the predicted age and chronological age.

We noticed that the model tends to overestimate younger subjects and underestimate older ones as shown in Fig.2d. This bias is previously reported as biased negative correlation between the brain age gap and the chronological age in other brain age studies [27, 28] and can be corrected using the linear correction. We tried the linear correction method, which helps to remove the bias, but it did not help to improve the prediction performance. As our study is not focusing on the brain age gap, we report results without correction. Details about the impact of the linear correction can be found in supplementary section 4.

A breakdown of subjects into roughly 5-year age groups (Fig.3) revealed that our models were the most effective for individuals aged 50-60, especially for people aged 55-60 years old, achieving the MAE of only 4.69 years.

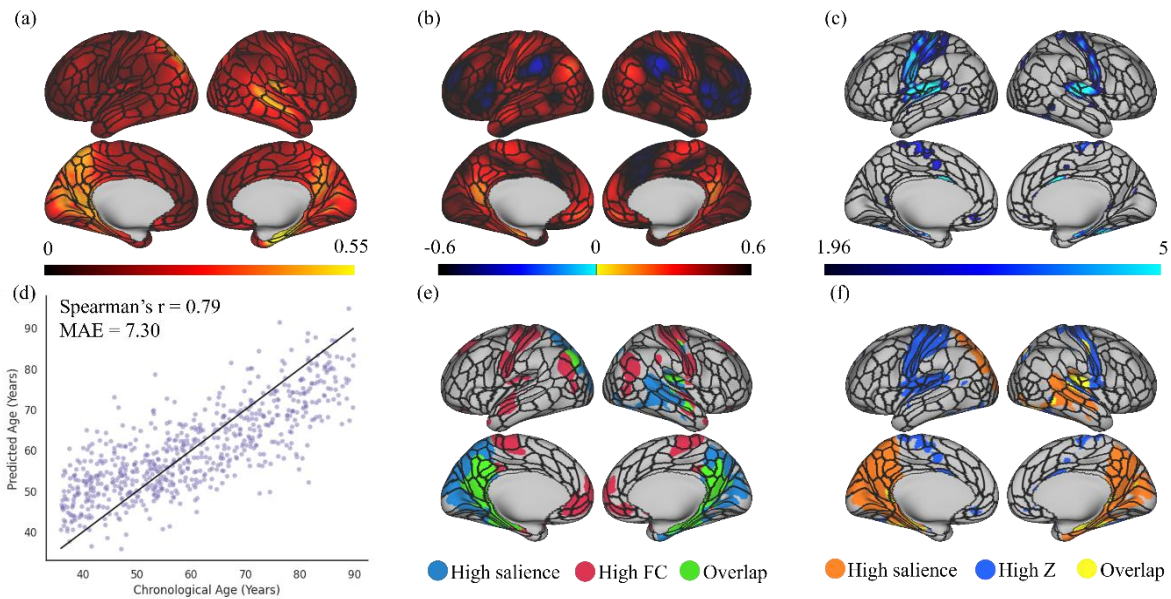

Fig. 2 (a) Average saliency map for age prediction using whole hippocampal cortical FC on the HCP-A dataset. The higher salience values (brighter colour) mean a higher contribution to the age prediction model. (b) Average hippocampal FC of HCP-A dataset. Positive values (red/yellow) are correlations, and negative (blue) are anticorrelations. (c) Thresholded z-scores derived from (Bonferroni-corrected) p-values obtained through linear regression analysis, showing regions with significant linear relationship between their hippocampal FC and chronological age. Z-scores are thresholded at 1.96. (d) The predicted age versus the chronological age for the HCP-A dataset. The black line shows the identity line (i.e. perfect prediction). (e) Overlapping map between (a) and (b). The saliency map and average FC map are thresholded to show the regions with the highest 20% values (highest absolute value for FC). (f) Overlapping map between (a) and (c). The saliency map is thresholded to show the regions with the highest 20% values. For (a-f), the MMP atlas ROI edges are overlaid on the brain surface to assist the identification of brain regions.

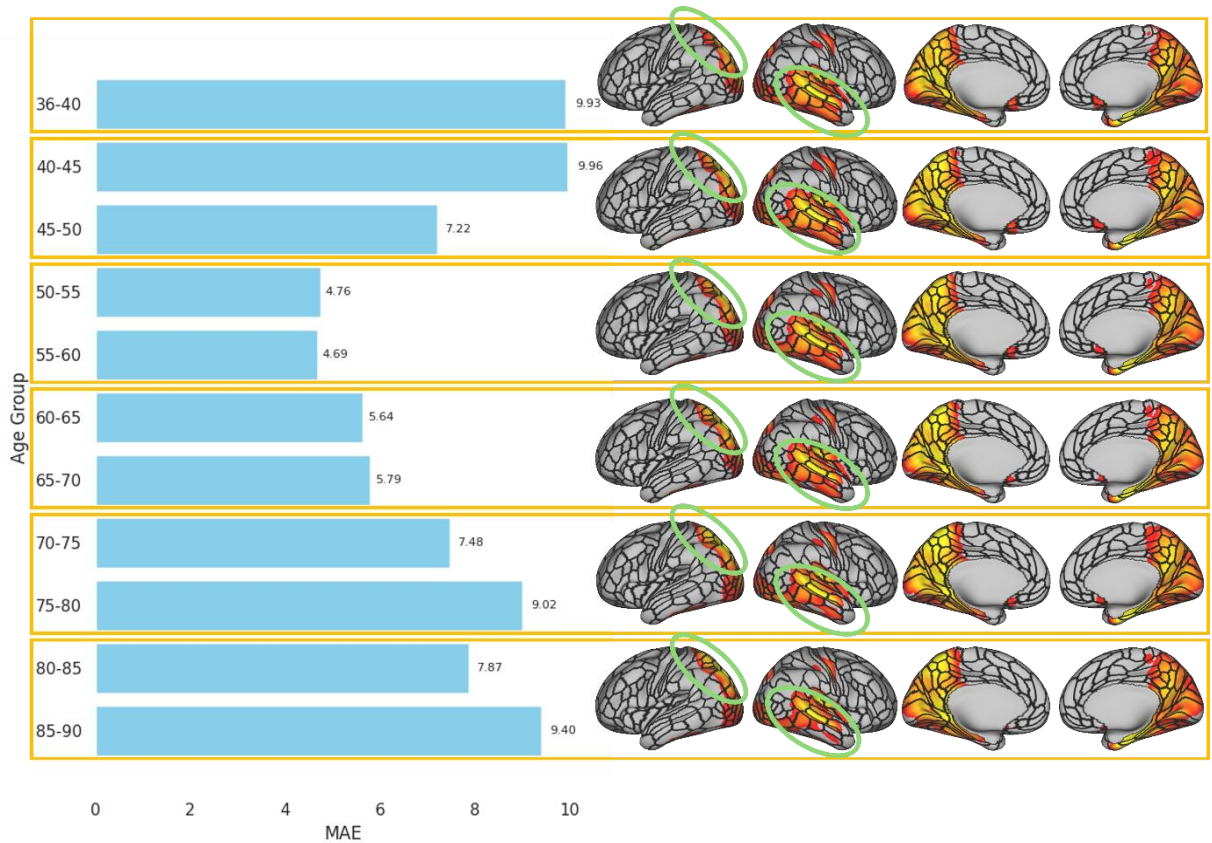

Fig. 3 Model prediction error and saliency maps for different age groups. Subjects were divided into smaller age groups to calculate the MAE. MAE values (in years) were denoted on the right side of each bar. Average saliency maps for each decade age groups are on the right side of this figure. Due to the subtle variations with age, saliency maps are shown for every decade range, for simplicity. From top to bottom, the saliency map corresponds to age groups: 36-40, 40-50, 50-60, 60-70, 70-80, and 80-90. Average saliency maps were thresholded at 0.33 (the threshold is set to be reserve top 20% values in the overall average saliency map). While overall similar, gradual changes (expansion or shrinkage) of the saliences identified in the right temporal lobe and the left superior parietal lobe are noticed and highlighted with green circles.

## 2.2. Interpretation of 3D CNN Models

To understand how our 3D CNN models made the prediction, we adapted LayerCAM to generate saliency maps; these maps visually represent the relative importance of different brain regions in predicting age based on their connectivity with the hippocampus.

With the LayerCAM, for each input FC and corresponding prediction, we generated one fused map from five convolutional layers for each subject. Subsequently, the saliency maps in the

3D volume space were mapped to the brain surface for better presentation and averaged across subjects, as shown in Fig.2a. In the saliency map, brighter regions in the saliency map indicate a greater influence on the model prediction. Based on the averaged saliency map, the hippocampal FC model emphasized connections between hippocampus and areas including PCC, PHC (inclusive of entorhinal cortex (EC)), left superior parietal lobule (SPL), right superior temporal sulcus (STS), and precuneus which are known to be affected in healthy aging and age-related diseases [9, 29].

For different age groups, we investigated the averaged saliency maps as shown in Fig.3. These saliency maps at different ages showed similar patterns with the overall averaged saliency map, indicating the stability of LayerCAM. Slight changes of the focus are found in the right inferior temporal and left superior parietal regions (highlighted by circles in Fig.3), i.e. the model can shift its focus to different parts of the brain for subjects at different ages, rather than focusing on the same spatial location for all ages. This finding may indicate that aging affects the hippocampal FC with these regions at different pace, and our model can capture such saliency at different age stage.

*Saliency maps vs. Average FC maps*

To investigate whether the saliency maps provide information beyond what it could have been inferred from the FC maps, we compared the two as shown in Fig.2(a-b). We thresholded the FC map and the saliency map to visualize the overlap between them as in Fig.2e. Thresholds for these maps were set to show brain areas holding the top 20% values.

According to the overlapping map, we identified that regions with relatively high FC value, such as the ventromedial PFC, are not necessarily highlighted in the saliency map. Conversely, regions like cuneus and right STS emphasized in the saliency map did not display high FC with hippocampus. These differences between the saliency maps and FC maps indicate that our model's interpretation method assesses the contribution of different brain regions beyond just their FC strengths, and the strong hippocampal FC regions may not be highly sensitive to aging.

*Saliency maps vs. General Linear Model Analysis*

To further explore the effectiveness of our method compared to general linear models, we performed a traditional linear regression analysis on FC maps, aiming to find significant linear age effect on the hippocampal FC. We mapped the z-scores ($z>1.96$, Bonferroni correction to the p-values) on brain surface to visualize regions linearly related to aging, with higher values indicating a more substantial linear association between hippocampal FC and chronological age. Similarly, we visualized the overlap between the saliency map and the z-score map (Fig.2f). Both the saliency map and the z-score map identified some regions in the PHC, but the saliency map further highlighted the precuneus, cuneus, PCC, and right STS, emphasising that our CNN model predictions provide more information beyond linear age effect.

To further validate the results, we did the linear and polynomial regression on regions identified by our 3D CNN model. All these regions show reduced FC with hippocampus in the aging process (Suppl. Fig. 5.). Our results show the simple regression is not strong enough to describe the relationship between hippocampal FC and age, however, adding second order or third order regressors improves the explainability. This result shows evidence that deep learning model can capture complex non-linear aging effect. More details can be found in supplementary information section 6.

## 2.3. Sex Difference

We also explored the differences between male and female subjects. In the HCP-A dataset, we have 313 males and 395 females. Our 3D CNN models achieved similar MAE of 7.23/7.35 years for the male/female, as shown in Fig.4a. The two sample t-test between the absolute errors of predictions on these two gender groups yielded a p-value of 0.76, showing no significant difference.

As shown in Fig.4b, the saliency maps are similar between these two sex groups. Both saliency maps identified the same regions as the overall average saliency map in Fig.2a. However, by doing the single-sided t-tests, we observed some differences between the saliency maps of these two sex groups as illustrated in Fig.4c. Specifically, the inferior parietal lobule (IPL), right middle temporal gyrus (MTG), precuneus, and part of the cuneus show relatively higher saliencies in female than in male. However, for male subjects, we did not observe significantly higher saliency than females. These results show our model not only captured core age-related hippocampal FC patterns across genders but also revealed subtle gender differences during normal aging.

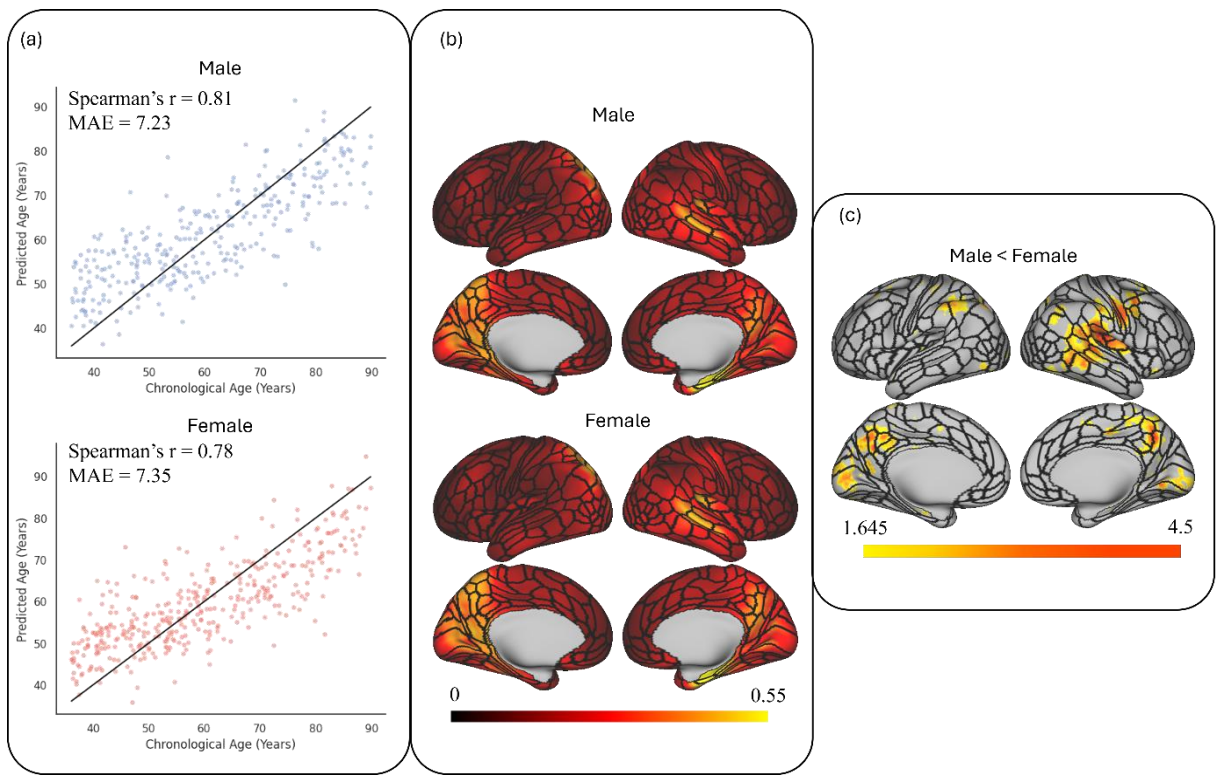

Fig. 4 (a) Age prediction made by the models using the whole hippocampal FC for female and male subjects respectively. The black lines are the perfect prediction. (b) Averaged saliency maps for female and male subjects. Higher salience values mean a higher contribution to the age prediction. (c) The FDR corrected t-tests' z-scores mapped to the brain surface, showing the differences between the saliency maps for male (M) and female (F) subjects. Z-scores were thresholded by 1.645. Hotter regions with higher z-scores indicate more significant differences.

## 2.4. The Anterior and Posterior Hippocampus FC

We trained another two models based on the anterior and posterior hippocampal cortical FC, yielding similar performance compared with the one using the whole-hippocampal cortical FC. After pretraining, the CNN predicted the age for young adults with an MAE of 3.27 years with the anterior hippocampal FC and 2.89 years with the posterior hippocampal FC. After fine tuning on the HCP-A dataset, the CNN using the anterior hippocampal FC predicted the chronological age with a validation MAE of 7.14 years. The one using the posterior hippocampal FC achieved the prediction with a validation MAE of 7.31 years. A one-way repeated measures analysis of variance (ANOVA) was conducted to statistically compare the performance across the three models using the whole, anterior, and posterior hippocampal cortical FC. The analysis yielded a p-value of 0.45 without correction and 0.43 after Greenhouse-Geisser correction, indicating there is no significant difference in MAE among the three models.

Average saliency maps (Fig.5b) for these two models using hippocampal subregions' FC identified similar regions to the saliency map from the whole hippocampal FC model (Fig.2a). However, when quantitively comparing the saliency maps from the anterior and posterior hippocampal cortical FC models, distinct patterns emerged. We quantified these differences using one-way paired t-tests and adjusted the resulting p-values using false discovery rate (FDR) correction and converted them into z-scores. These differences are visualized on the brain surface with a threshold of $z > 1.65$ in Fig.5c, where higher z-scores indicate more pronounced disparities between the anterior and posterior hippocampal cortical FC models. Specifically, FC between the anterior hippocampus and regions including the PFC, anterior cingulate cortex

(ACC), EC, and temporal pole (TP) were more influential on age prediction compared with their posterior hippocampal connections. Conversely, FC between the posterior hippocampus and precuneus, IPL and STS had a stronger predictive influence than their FC with the anterior hippocampus. These differences reveal the functional segregation of the anterior and posterior hippocampus, as well as the sensitivity of their cortical FC to age.

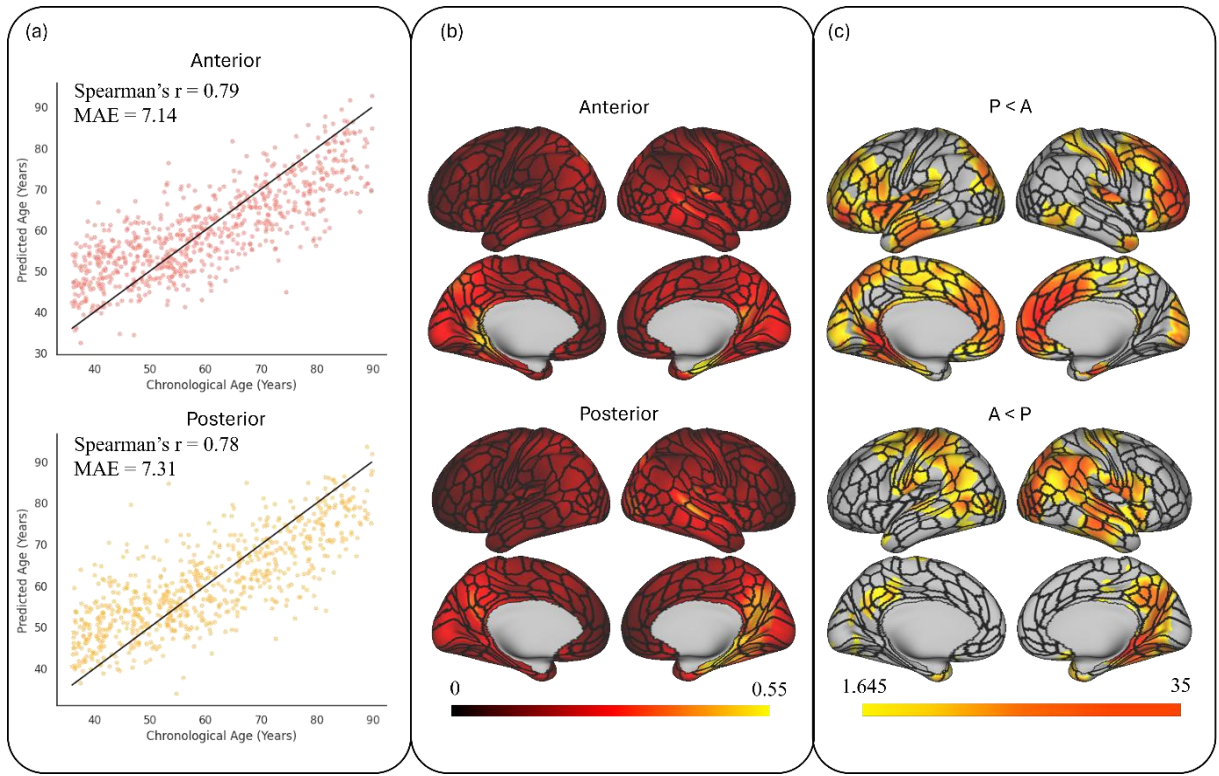

Fig. 5 (a) Age prediction made by the models using the anterior and posterior hippocampal FC, separately. The black lines are the perfect prediction. (b) Averaged saliency maps based on models using anterior and posterior hippocampal FC. The higher salience values mean a higher contribution to the age prediction. (c) The corrected t-tests' z-score mapped to the brain surface, showing the differences between the saliency maps based on anterior (A) and posterior (P) hippocampal FC. Z-scores were thresholded by 1.645. Hotter regions with higher z-scores indicate more significant differences.

## 3. Discussion

We presented here an interpretable 3D CNN model of brain age prediction based on the voxel-wise FC to the hippocampus. By generating saliency maps, we identified brain regions whose hippocampal FC is highly relevant to the normal aging process. Comparisons between gender groups and hippocampal subregions revealed significant differences. Our findings also indicate that hippocampal FC can predict age with an accuracy comparable to existing methods. Importantly, our voxel-level model offers enhanced spatial details and interpretability, uncovering localized aging patterns often overlooked by region-based approaches.

### 3.1. Functional Connectivity Between the Whole Hippocampus and Cortex as an Age Predictor

We used LayerCAM to generate saliency maps, representing a novel approach to enhancing the interpretability of deep learning models in neuroimaging context and facilitating the acceptability and utility of model predictions. Unlike methods such as the gradient-weighted class activation mapping (Grad-CAM), LayerCAM integrates the activation of multiple convolutional layers, producing saliency maps with higher spatial resolution, a crucial advantage for neuroimaging studies requiring fine-grained localization. A comparison between Grad-CAM and LayerCAM is provided in supplementary section 5.

Our saliency maps generated using LayerCAM revealed how voxel-wise hippocampal-cortical FC contributes to chronological age prediction, supporting the biological relevance of the model's predictions. Our discussion would mainly focus on the regions identified, different saliencies between two brain hemispheres, and different saliencies between the two sex groups.

*Age Sensitive Hippocampal-Cortical FC*

Saliency maps highlighted hippocampal FC with regions including precuneus, cuneus, PCC, PHC, left SPL, and right STS to be highly significant in predicting age (Fig.2a). These regions are functionally linked to visuospatial and cognitive processing which are sensitive to age-related changes. For instance, some default mode network (DMN) regions (precuneus, cuneus, PCC, and PHC) play central rules in different elements of visuospatial related cognitive functions [25, 30-33], while the SPL is involved in sensory integration and spatial orientation [34, 35].

Our findings are consistent with prior studies reporting age-related structural and morphological changes. Specifically, the precuneus and cuneus showed cortical thinning [36, 37], and the STS exhibits widening and reduced depth in older adults [38]. Functionally, the DMN (including the precuneus, PCC, and PHC) demonstrates reduced FC with increasing age [39, 40], and reduced cerebral blood flow in the SPL has also been reported during rest [41]. Additionally, increased task-related activation in the cuneus among older adults during visual problem-solving tasks has been interpreted as evidence for functional compensation [42]. Our results further support the resting-state functional relevance of this region in aging.

These results show that our 3D CNN model captured age-sensitive, biologically plausible FC patterns. Our results extend previous research by identifying fine-grained voxel-level hippocampal-cortical interactions that may serve as functional markers of normal aging. These hippocampal FC may be further explored in patients with neurodegenerative conditions, which could aid in distinguishing normal aging from early neurodegenerative changes in clinical settings.

*Patterns in Saliency Maps Reveal Asymmetry in Aging-Related FC*

We observed hemispheric asymmetries in the saliency map, specifically in the left SPL and right STS. These results extend prior observations of stronger left SPL activation in older adults during episodic memory retrieval task [43], by showing that even during a resting state, hippocampal connectivity with left SPL is more relevant to aging compared to right SPL. Our findings also align with the reported morphological changes that the right STS fold opening increases during aging, which is faster than the left STS [44]. The functional asymmetries detected by our model mirror known neuroscience knowledge.

Additionally, in our saliency map, bilateral PHC is highlighted but emphasized on the right side, especially it extends to right EC. Prior studies found both left and right EC experiences age-related volume reduction, but the right EC shows a higher atrophy rate compared to the left [45-47]. Therefore, the heightened salience of right EC connectivity may reflect its greater vulnerability to aging and its potential value as a biomarker for age prediction.

Collectively, these patterns suggest the model is sensitive to hemisphere-specific FC changes associated with normal aging, providing insights into the differential role of detailed regions in the aging brain.

*Saliency Maps Reveal Sex Differences in Aging Process*

While average saliency maps for males and females showed similar patterns, indicating the model capture stable and core hippocampal FC features across sexes, we detected subtle sex differences via t-tests (Fig.4c). Specifically, the model showed greater salience in certain regions, including precuneus, for females compared to males, suggesting higher age-sensitivity of FC in those areas.

Existing literature reported sex differences in hippocampal FC [48], and females are generally found to be more vulnerable to cognitive decline and Alzheimer's disease [49-51]. Our results indicate that hippocampal FC features may reflect these sex-specific susceptibilities, and that deep learning models can detect these nuances without explicit sex labels, purely depending on the FC impact to age prediction. Moreover, these findings underscore the importance of considering sex as a biological variable in brain aging studies.

### 3.2. Distinct Contributions of Anterior and Posterior Hippocampus to Age Prediction

Despite similar model performance across whole, anterior, and posterior hippocampal FC inputs, the saliency map comparison revealed distinct age-predictive FC patterns for hippocampus subregions. This suggests that while both the anterior and posterior hippocampus provide valuable information for age prediction, they do so via functionally different cortical regions, highlighting the richness of patterns in hippocampal FC and the robustness of our methods.

*Anterior Hippocampus*

Compared to the posterior hippocampus, age-predictive salience for the anterior hippocampus was stronger in its FC with the PFC, ACC, EC, and TP, which are involved in cognitive functions that are vulnerable to aging, including emotional regulation and decision-making [52-54]. The ventromedial PFC (vmPFC) is known to connect with the anterior hippocampus [55], playing a key role in emotion regularization [56]. Moreover, the anterior hippocampus is more anatomically [18] and functionally connected to areas within the TP, especially the right TP [57]. Our findings extend these observations, emphasizing that anterior hippocampal FC with PFC and TP is highly sensitive for predicting age, potentially reflecting aspects of age-related cognitive decline.

Additionally, the EC, a key interface between the hippocampus and neocortex [58], communicates between both the anterior and posterior hippocampus and the cortex [55]. Our results suggest the anterior hippocampus-EC pathway may be particularly sensitive to age-related changes, suggesting that this anterior EC pathway may serve as a marker for healthy aging. Although previous work [55, 59] indicates functional connections between the posterior hippocampus and ACC, our model is more focused on FC between it and the anterior hippocampus, possibly reflecting a shift in functional dominance as aging progresses.

*Posterior Hippocampus*

In contrast, the posterior hippocampal showed stronger salience in its FC with the precuneus, IPL, and STS for age prediction. These regions are involved in episodic memory retrieval [30], spatial attention [60], and social cognition [25, 60, 61], showing their high relevance to age-related cognitive decline.

Our findings align with previous rs-fMRI studies that found both precuneus and bilateral IPL were more correlated to the posterior hippocampus than the anterior hippocampus [59]. During aging, the FC between the posterior hippocampus and IPL was found to decrease [62]. Our results support and extend these findings by showing that posterior hippocampal FC with the precuneus and IPL is predictive of age and may reflect early functional changes associated with normal aging. Given that IPL is also affected in AD [63], this FC may represent an early sign of age-related cognitive decline. Interestingly, while prior research emphasized anterior hippocampal FC with the lateral temporal cortex and TP [55], we observed age-sensitive FC between the posterior hippocampus and superior temporal regions, broadening our understanding of temporal lobe contributions to aging.

Collectively, these results highlight that anterior and posterior hippocampal connectivity patterns reflect distinct functional trajectories in normal aging, which is potentially linked to the heterogeneity of age-related cognitive decline. The detailed differential aging effects of the anterior and posterior hippocampus FC we revealed may establish new directions to develop a more comprehensive understanding of the human brain functional changes during normal aging.

### 3.3. Model Performance in Predicting Age

Our model achieved the MAE of 2.99 years for young adults in the HCP-YA dataset, comparable to previous work using structural or diffusion MRI on the same dataset (MAE: 2.44 - 2.59 years) [64-66]. For the HCP-A dataset, our MAE increased to 7.30 years using the whole hippocampal cortical FC. Although a recent study [67] combining task fMRI, resting-state fMRI, and structural MRI achieved an MAE below 5.4 years, our model outperformed their resting-state-only model (MAE: 7.87 years). Another study using whole-brain FC matrices from the UK Biobank database of 10,065 subjects with a narrower age range (40 to 69 years) reported an MAE of 4.82 years [68]. In comparison, our model achieved an MAE of 6.47 years in the similar age range between 40 to 69 years, narrowing the performance gap to just 1.65 years but with finer-grained FC information.

Notably, our application of the model across a wider age range can introduce more variations from the data, making it more challenging for age prediction. Moreover, we demonstrated the age prediction power of hippocampal FC on the whole-brain voxel level is approaching the age prediction based on regional or network level analysis. Importantly, the use of voxel-wise analysis provides finer-grained mapping of age-related FC changes. This study, therefore, demonstrates the efficacy of deep learning in neuroimaging and paves the way for more targeted focus of brain regions related to normal aging.

Our model's predictions were more accurate for middle-aged individuals. This aligns with evidence of accelerated cognitive changes after middle age [69, 70], which may be associated with more individual variances complicating the age prediction in the elderly. At younger ages, although some cognitive functions alter [71], changes in the hippocampal FC before 50 years may be quite subtle to detect and use for age prediction, leading to higher prediction error than in middle ages.

### 3.4. Additional Technical Insights

The comparison between our interpretable deep learning method and conventional regression analysis underscores the complexity of brain aging and the intricacies of hippocampal FC's role in this process. While traditional linear regression on hippocampal-cortical FC revealed the existence of linear correlations between age and hippocampal FC, particularly in the motor cortex, somatosensory cortex, superior temporal gyrus (STG), and partial PHC, limitations of this linear model became apparent when our 3D CNN model uncovers more regions with complex and nonlinear patterns in their hippocampal FC, such as precuneus and PCC.

The linear regression highlighted the motor cortex, somatosensory cortex, and STG as regions with the most substantial linear relationship with age, as indicated by their high z-score values. Although the functions of these regions may also be affected during aging, they are not classically associated with cognitive decline. Importantly, the low $R^2$ values across regions indicate the weak explanatory power of linear regression and high individual variability in the hippocampal FC (Supplementary Section 6). These discrepancies suggest that the linear model is not sufficient at capturing the intricate changes of the relationship between hippocampal FC and normal aging, limiting its utility for extracting more subtle neural dynamics from large individual variance.

Additionally, linear regression operates on a vertex-wise basis, neglecting spatial dependencies. By incorporating neighbouring voxels, our 3D CNN can capture spatial patterns of hippocampal FC that better reflect complex brain functional organization, enhancing its sensitivity and accuracy in capturing complex and spatially distributed brain FC pattern changes during aging. These findings support the use of interpretable deep learning as the method for uncovering complex FC-aging relationships. By combining predictive power with transparency, our model offers new opportunities to explore individual aging trajectories and informs preventative strategies or interventions.

### 3.5. Limitations and Future Directions

Our 3D CNN model enhances interpretability and is sensitive to distinct subregions of hippocampal FC. However, further investigation is required to assess its potential for clinical applications. Exploring the brain age gap, i.e. the difference between predicted and chronological age, allows for individualized analysis. Individuals with a younger predicted brain age than their chronological age may possess neural characteristics associated with a "younger" brain, whereas those with overestimated brain ages may reflect an "older" brain. The brain age gap as the difference between brain age and chronological age could serve as a baseline for evaluating brain age [28, 72]. Moreover, correlating the brain age gap with cognitive measures, behavioural data, genetic information, etc., offers novel insights into prediction errors and considers various factors affecting brain aging [28, 73], benefiting precision medicine and personalized treatment.

Our current focus on the hippocampus for FC calculation also provides limited perspective. Exploring alternative ROIs or incorporating additional relevant subcortical and cortical regions could potentially enhance predictive accuracy. This could also provide a more comprehensive understanding of the human brain functional changes during the aging process.

Currently, our analysis is confined to a healthy aging cohort. Applying our model, trained on healthy cohort, to populations with neurological conditions like dementia could offer benchmarks for evaluating deviations from typical aging trajectories. Furthermore, expanding the model's focus across the entire lifespan presents a compelling, but challenging, prospect, especially when including infants and children, due to the substantial variability in brain development stages.

## 4. Conclusion

The hippocampus is affected during normal aging process, but its cortical FC relationship with the aging is not fully explored. In our study, we employed a 3D CNN model to predict age based on voxel-wise cortico-hippocampal FC and adapted LayerCAM for interpretative analysis. The use of LayerCAM enables the transparency of our deep learning method. The regions identified as most contributory to age prediction, such as precuneus, PCC, and cuneus, align with those known to undergo changes during healthy aging. Through a finer-grain analysis by splitting the hippocampus into the anterior and posterior subregions, we further validated our methods and uncovered differences in the contributions of anterior and posterior hippocampal cortical FC to the age prediction task, indicating our method's sensitivity in detecting FC variations along the hippocampus's long axis during the aging process. Many of the regions identified in our study were not detectable using linear regression, proving the efficacy of our 3D CNN. Our study offers a novel, interpretable deep learning approach to uncover complex hippocampal FC changes in healthy aging, which has the potential to advance the understanding of age-related neural connectivity and guide future interventions to support cognitive health in older adults.

# 5. Methods

## 5.1. Dataset and Preprocessing

The publicly available HCP-Y [23] and HCP-A [24] datasets were used in this study. These two datasets have some differences in their imaging protocols and the age distribution. The rs-fMRI in HCP-Y has a spatial resolution of 2mm isotropic, TR = 720ms, and TE = 33.1ms, while the HCP-A has a spatial resolution of 2mm isotropic, TR = 800ms, and TE = 37ms. We used the minimally preprocessed [74] rs-fMRI from both datasets. In both datasets, each participant has four resting-state fMRI scans. Each HCP-Y rs-fMRI run contains 1200 frames, but each of the HCP-A rs-fMRI runs only records 488 frames. All these frames were utilized to calculate the FC. Excluding subjects with missing scans, we used volume-based fMRI data from 1018 subjects, aged from 22 to 37 with a mean of 28.7, of the HCP-Y dataset. Excluding subjects with missing scans and those aged above 90 years old without detailed age information, we included 708 subjects from HCP-A give a wider age range from 36 to 90 years (age and gender distribution can be found in supplementary section 1).

Both data sets were minimally preprocessed while being released. To further remove noise in both spatial and temporal domains for the rs-fMRI data from HCP-Y and HCP-A, we applied spatial smoothing with a 6mm full width at half maximum, followed by a bandpass filter with 0.01 to 0.1 Hz to filter out noises in the temporal domain. Because our scope is on normal aging rather than healthy young adults, the HCP-Y dataset is only used for model pretraining, and the main results are based on the fine-tuned models on the HCP-A dataset.

### 5.2. Seed-Based Functional Connectivity

We calculated two sets of seed-based FC by setting the seed regions in the left and right hippocampus. The individualized hippocampus mask was provided with the minimally preprocessed HCP datasets. We extracted and averaged the time series across voxels in the left and right hippocampus, respectively. The averaged time series for each of the ROIs was used to calculate the whole-brain voxel-wise FC using Pearson's correlation. For each individual, the FCs calculated from four rs-fMRI scans were averaged to generate the final individual hippocampal cortical FC for both datasets.

### 5.3. 3D Convolutional Neural Network

The model we used to predict the chronological age is a 3D CNN that can perform 3D convolution on the input voxel-wise FC data. The model architecture is illustrated in Fig.1a. The input to the CNN is the hippocampal cortical FC for each individual. The FC based on the left and right hippocampus were treated as two input channels to the model. Specifically, the model is composed of stacking five convolutional steps. In each convolutional step, the input is first passed to the 3D convolutional layer, after which it undergoes batch normalization in 3D space. The output of batch normalization was applied to the ReLU activation and then passed to the max pooling layer with the kernel size of 2x2x2 and the stride of 2. Inspired by

Inception models [26], we utilized convolution kernels with different sizes to capture both local and global information from the FC data. Two types of convolutional layers, using the convolution kernel of 3x3x3 or 5x5x5, were applied alternatively. We used different sizes of the kernel because small kernels, like the one with size 3x3x3, can capture fine-grained local features, while large ones, like the 5x5x5 kernel, can get coarse global features from the data. All convolutional layers used in this model used a stride of 1x1x1 with padding. The numbers of output channels were set to 16, 16, 32, 32, and 64, respectively, for convolutional layers from shallow to deep. After five convolutional processes, the output from the max-pooling layer of the last block was flattened and passed to a dropout layer followed by a linear layer to directly output the predicted age. The performance of the model was compared to other state-of-the-art CNN architectures adapted for our data and task for model selection purpose. More information can be found in supplementary section 3.

### 5.4. Pretraining and Fine Tuning

Although we are focusing on the healthy aging cohort, we utilized the HCP-Y to pretrain our model. The main aim is to improve the model stability and reduce the risk of overfitting. With a more focused age range, the young adult dataset may have less noise and more consistent patterns of FC compared to the aging dataset. Thus, pretraining has the potential to make the model learn a robust set of features and provide a more stable initial weight configuration. Additionally, pretraining on the HCP-Y dataset allows the model to start with a broad understanding of FC, reducing the need to learn from scratch on the smaller aging dataset.

For pretraining, we split the data into training and testing sets with a ratio of 4:1, so we have 814 young adults for training and 204 individuals for testing. We trained a 3D CNN model to predict age with the same architecture to minimize the MAE using the whole hippocampus cortical FC, respectively, where

$$MAE = \frac{\sum_{i=1}^{n} |y_i - \hat{y}_i|}{n} \qquad \text{Eq. 1}$$

$y_i$ is the true chronological age, $\hat{y}_i$ is the predicted age, and $n$ is the number of subjects. The initial learning rate was set to 0.00002 and Adam was used as the optimizer. These models were trained for 50 epochs and a batch size of 16. These three models were then evaluated on the test dataset.

After pretraining on the larger HCP-Y dataset, we fine-tuned our models on the HCP-A dataset for domain adaptation. Overfitting was mitigated through the application of fivefold cross-validation and early stopping. The dataset was divided into five non-overlapping subsets, each containing 144 subjects. In each fold, one subset of the data was used for validation, and the other 576 subjects' FC were used for training. The training and validation loss were monitored and recorded. The patience of early stopping was 15 and the maximum number of epochs was 200, which means the training and validation will stop if the validation performance is not improved for more than 15 epochs or it reaches the maximum number of training epochs.

The implementation of the model was done through PyTorch. The model was trained and validated with NVIDIA A100-PCIE-40GB GPUs (driver version: 450.248.02, CUDA version: 11.0).

### 5.5. Deep Learning Interpretation

For classification problems using CNN, CAM [75] can generate saliency maps that highlight which part of the input contributed more to the prediction. In this category, Grad-CAM [76] is a widely used one. Grad-CAM use the weighted feature map of the last convolutional layer to generate saliency maps, highlighting the contribution of different parts of the input to the specific class prediction. As a variation of Grad-CAM, LayerCAM [17] fuses the saliency maps generated at each convolutional layer, yielding much higher spatial resolution. Thus, this method was adapted to generate saliency maps for our age prediction to help explain the model behaviour. With $g_{xyz}^{kc}$ as the gradients at location (x, y, z) for k-th feature map at a specific layer, LayerCAM weights feature maps (A) using the positive portion of the gradients as in Eq. 2, so that each location will have a corresponding weight $w_{xyz}^k$, rather than having the channel-wise weight. Finally, LayerCAM generates the saliency map (M) by linearly combining the weighted feature maps $\hat{A}^k$ along the channel dimension as formulated in Eq. 4.

$$w_{xyz}^k = relu(g_{xyz}^k) \qquad \text{Eq. 2}$$

$$\hat{A}_{xyz}^k = w_{xyz}^k * A_{xyz}^k \qquad \text{Eq. 3}$$

$$M = ReLU(\sum_k \hat{A}^k) \qquad \text{Eq. 4}$$

After getting the saliency maps from each convolutional layer, we scaled the saliency maps generated by shallower convolutional layers (except for the last convolutional layer) in the same way proposed in the original study [17], because shallower layers have lower activation values. The scaling process is defined in Eq.5 using a scaling factor $\gamma$ and $L$ denotes the total number of convolutional layers. As Jiang et al. [17] found $\gamma = 2$ yielded the best localization ability, this value was used in our study. Although the original LayerCAM use the element-wise maximum to generate the combined map, we combined scaled saliency maps from all convolutional layers through simple summation and normalization to preserve the contribution of all layers and avoid the domination of potential extreme values in saliency map from deeper layers, generating a comprehensive and stable map.

$$\widehat{M}_i = \tanh\left(\frac{\gamma * M_i}{\max(M_i)}\right), i = 1, 2, \ldots, L - 1 \qquad \text{Eq. 5}$$

$$M_{fused} = \frac{M_n + \sum_i^{L-1} \widehat{M}_i}{\max(M_n + \sum_i^{L-1} \widehat{M}_i)} \qquad \text{Eq. 6}$$

Although CAM methods are commonly used in classification problems, we adapted it to our regression tasks, in which we try to predict continuous quantitative scores or values. For classification problems, CAM highlights the intensity changes in certain input pixels (or voxels in 3D space) that have the most impact on the prediction score [76], leading to a specific category. In our regression task, we omitted the step to transfer the predicted score to a specific class label. Thus, the saliency maps generated from LayerCAM would highlight the regions that contributed the most to making the specific age prediction as a continuous number.

After getting saliency maps based on each individual's prediction, we calculated the group average saliency map across all subjects. To investigate whether brain regions with higher FC leads to higher saliences in the maps, we averaged FCs across all subjects and compared them with the average saliency map. The FC map and the saliency map were thresholded to show brain locations holding the top 20% values. Specifically, the FC map was thresholded to retain brain location with the absolute value of FC higher than 0.21, and the saliency map used the threshold of 0.39.

### 5.6. Linear Regression Analysis

Linear regression is a widely used traditional method to study age effect on human brain [8, 77]. For comparison purpose to the 3D CNN method proposed here, we also performed linear regression on 708 subjects, excluding subjects aged 100-year-old. The chronological age was used as the regressor to fit the hippocampal cortical FC map generated from each hippocampal seed-ROI analysis (see Section 5.2). For each subject, the averaged hippocampal FC map was obtained by averaging the left and right hippocampal FC maps. The linear regression was applied to the hippocampal FC mapped to the brain surface, which is defined as below:

$$FC = \beta Age + \epsilon \qquad \text{Eq. 7}$$

where FC is the hippocampal cortical FC on brain surface, $\beta$ is the coefficient, $\epsilon$ is the error, and age is the chronological age. We have multiple linear models on surface vertex level and obtained the p-value and $R^2$ to investigate the significance and strength of the linear relationship. P-values were corrected using the Bonferroni correction and converted to z-scores. We thresholded the z-score at 1.96 and visualized on the brain surface for visual inspection.

To further investigate the regions identified in the saliency maps, we generated masks for precuneus, cuneus, PCC, PHC, left SPL, and right STS based on the MMP atlas as shown in Suppl. Fig.5. These masks were used to extract the regional hippocampal FC for each subject. For these regional hippocampal FC, we performed linear, second-order polynomial, and third-order polynomial regression. For all the regression models. We recorded $R^2$ to explore the strength of their hippocampal FC relationship with age.

### 5.7. Sex Group Comparison

Based on the 3D CNN using the hippocampal FC for age prediction, we further explored the sex group differences. Among the 708 subjects (excluding 12 subjects aged 100 years), we have 313 male subjects and 395 female subjects. The MAE was calculated for these two groups. For a more thorough comparison of the model's prediction performance on these two groups, we employed the two-sample t-tests to compare the absolute error in their age prediction.

Aside from the performance, we also investigated their saliency maps to check the contribution of the hippocampal FC to the age prediction. For each group, we averaged the saliency maps across subjects. We would expect to find some differences between these two gender groups. Thus, we carried out two one-way t-tests. One of them has the alternative assumption that the mean of the distribution underlying the saliency maps for the male is less than the mean of the distribution underlying saliency maps for the female. The other one has the inverse alternative assumption. The resulting p-values on voxel-level were corrected through FDR correction and converted to z-scores. For visualisation purposes, z-scores were thresholded at 1.65 and mapped to the brain surface.

## 5.8. Anterior and Posterior Hippocampal FC Analysis

The hippocampus is heterogeneous, with the anterior and posterior parts known to have different functional roles. Specifically, the anterior hippocampus is more relevant to memory encoding and emotion regulation, while the posterior hippocampus is responsible for memory retrieval and navigation [19-22, 78]. Given normal aging may affect them differently, and to test our method's specificity, we further divided the hippocampus into anterior and posterior regions.

To generate the mask for anterior and posterior hippocampus, an experienced neuropsychologist first manually separated the anterior and posterior hippocampus from the subcortical ROIs atlas provided in HCP dataset (see Suppl. Fig.2). By reverse registering this segmentation of anterior and posterior hippocampus to individual hippocampus masks, we only found minor misalignment. Therefore, we can reliably divide the individual hippocampal mask into anterior and posterior parts on both hemispheres. With these four seeds, i.e. left posterior, right posterior, left anterior, and right anterior hippocampus, we calculated their FC to the whole brain for each subject.

We then used the same methods in 5.3-5.5 to generate two 3D CNN models separately for the anterior and posterior hippocampal FC. The validation prediction errors of these two models plus the one working on the whole hippocampal FC were compared using a one-way repeated measures ANOVA test.

As for the saliency maps based on the models computed using anterior or posterior hippocampal cortical FC maps, we would expect to capture some differences between the two sets. To compare these two sets of saliency maps, we performed two one-way paired t-tests. One of them has the alternative assumption that the mean of the distribution underlying the saliency maps based on anterior hippocampal cortical FC is less than the mean of the distribution underlying saliency maps based on the posterior hippocampal cortical FC. The other one has the inverse alternative assumption. The resulting p-values were corrected by the FDR correction and converted to z-scores. For visualisation purposes, z-scores were thresholded at 1.65 and mapped to the brain surface.

# 6. Acknowledgement

Data used in this publication was supported by the National Institute on Aging of the National Institutes of Health under Award Number U01AG052564 and by funds provided by the McDonnell Center for Systems Neuroscience at Washington University in St. Louis. The HCP-Aging 2.0 Release data used in this report came from DOI: [10.15154/1520707](10.15154/1520707). JL is supported by Brain and Mind Centre Research Development Grant, USYD-Fudan Brain and Intelligence Science Alliance Flagship Research Program, Moyira Elizabeth Vine Fund for Research into Schizophrenia Program. JL, FC and SN are supported by an ARC Discovery Project (DP240102161).

# Supplementary Information

## 1. Age Distribution of the Dataset

The age distribution of the human connectome project aging (HCP-A) dataset is illustrated in Suppl. Fig.1. Most of the participants are aged between 35 to 60, while there are relatively fewer participants aged from 75-90.

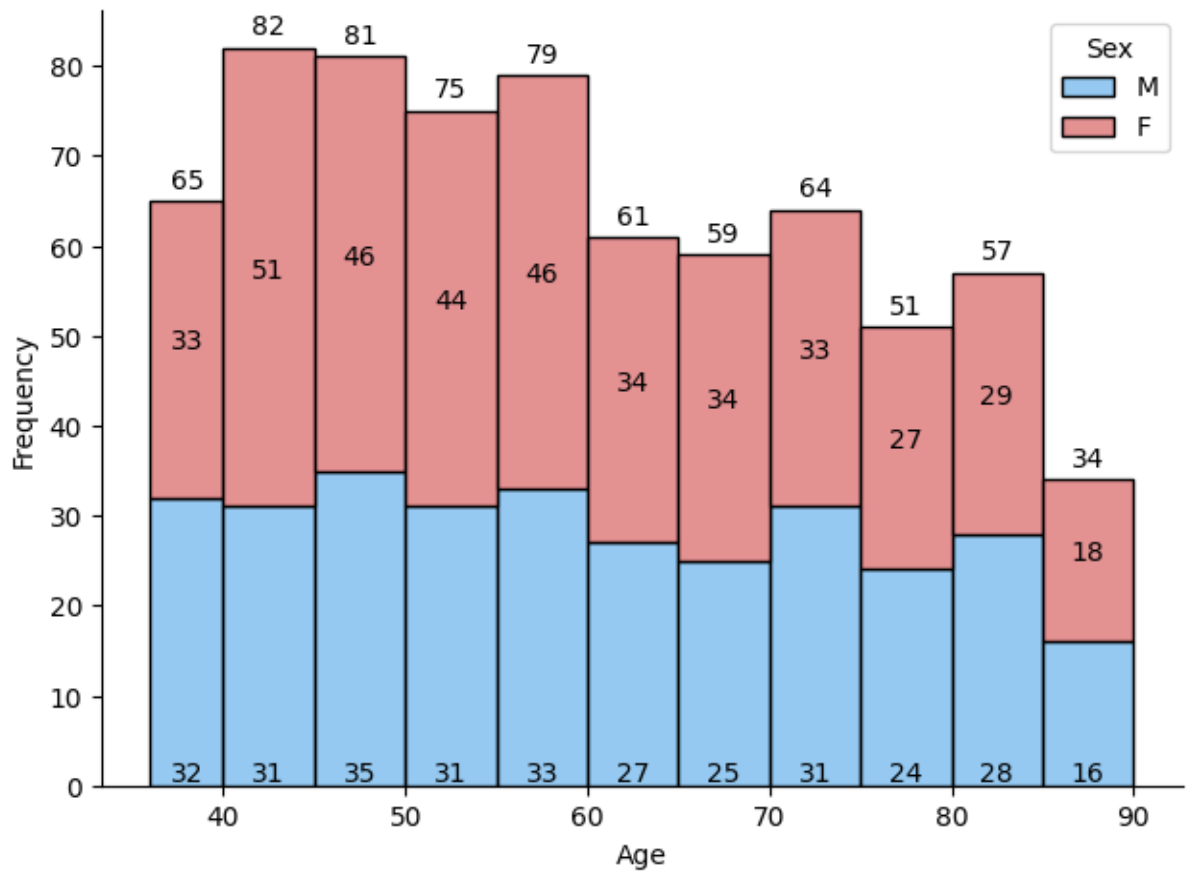

**Suppl. Fig. 1.** Age Distribution by gender. Blue bars stand for the number of males and the pink bars are for the number of females. The number within each bar is the count for either males or females. The number on top of each bar is the total number of subjects in different bins disregarding the sex (equal to the sum of two numbers within the bar).

## 2. Anterior and Posterior Hippocampus Masks

The anterior and posterior hippocampal masks we used for the seed-based FC analysis are shown in Suppl. Fig. 2. We first obtained the hippocampal mask from the atlas of subcortical regions provide with HCP minimally preprocessed data. The hippocampus was then manually split into the anterior and posterior parts by experienced neuroscientists.

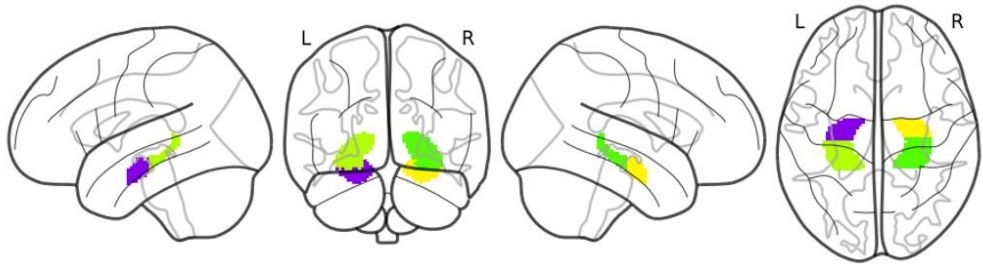

**Suppl. Fig. 2.** The atlas for the anterior and posterior hippocampus in the MNI space. Bilateral hippocampus were divided into the anterior and posterior shown in different colours.

## 3. Model Selection of CNN Architectures

Purely based on the HCP-A dataset, we compared the performance of different models on the age prediction task, including some popular and commonly used models in computer vision, such as VGG [1], ResNet [2] and EfficientNet [3]. We implemented and adapted VGG13, ResNet12, ResNet26, ResNet50, EfficientNet, and SFCN [4] for predicting age using the whole-hippocampal cortical FC data as 3D volumes. The performance was evaluated based on the fivefold cross validation on the HCP-A dataset. No pretraining was used here for the model comparison. The performances, quantified using MAE, are listed in Suppl. Table 1. Our 3D CNN model outperformed other existing models.

**Suppl. Table 1.** Performance of different 3D CNN models on HCP-A dataset. This table contains the mean absolute errors (MAEs) for different 3D CNN models based on the validation dataset in the fivefold cross-validation. The 3D CNN model introduced in Section 5.3 showed the best performance with less than 10 years of prediction error without pretraining.

| Model Name | Validation MAE (Years) |
|---|---|
| VGG13 | 10.29 |
| ResNet12 | 22.82 |
| ResNet26 | 13.83 |
| ResNet50 | 17.70 |
| EfficientNet | 14.62 |
| SFCN | 21.37 |
| **Our 3D CNN model** | **9.82** |

## 4. Age Prediction Linear Bias Correction

As noticed, our model tends to overestimate the age of young subjects and underestimate the age for elder ones, showing a negative correlation between the brain age gap and the chronological age as illustrated in Suppl. Fig.3 (b). We applied the linear correction method commonly used in other studies [5, 6]. To correct the bias, we first fit a linear line between the predicted age (y) and the chronological age (x) as specified in Eq.1. Using the estimated parameters, we calculated the corrected age gap using Eq. 2.

$$y = \alpha x + \beta \quad \text{Eq. 1}$$

$$Corrected\ age\ gap = \frac{y - \beta}{\alpha} - x \quad \text{Eq. 2}$$

Although other works reported the improved performance after the correction, we found the variance of the corrected age gap (123.34) become much larger than the uncorrected age gap (80.75) and the MAE also increased a little bit (from 7.30 to 8.79). Suppl. Fig.3 visualized the predictions before and after correction, showing the bias was successfully removed. In this study, we are not focusing on the brain age gap and its relationship with other behavioural or cognitive measurements. Instead, we value the age prediction accuracy, so we report the uncorrected results.

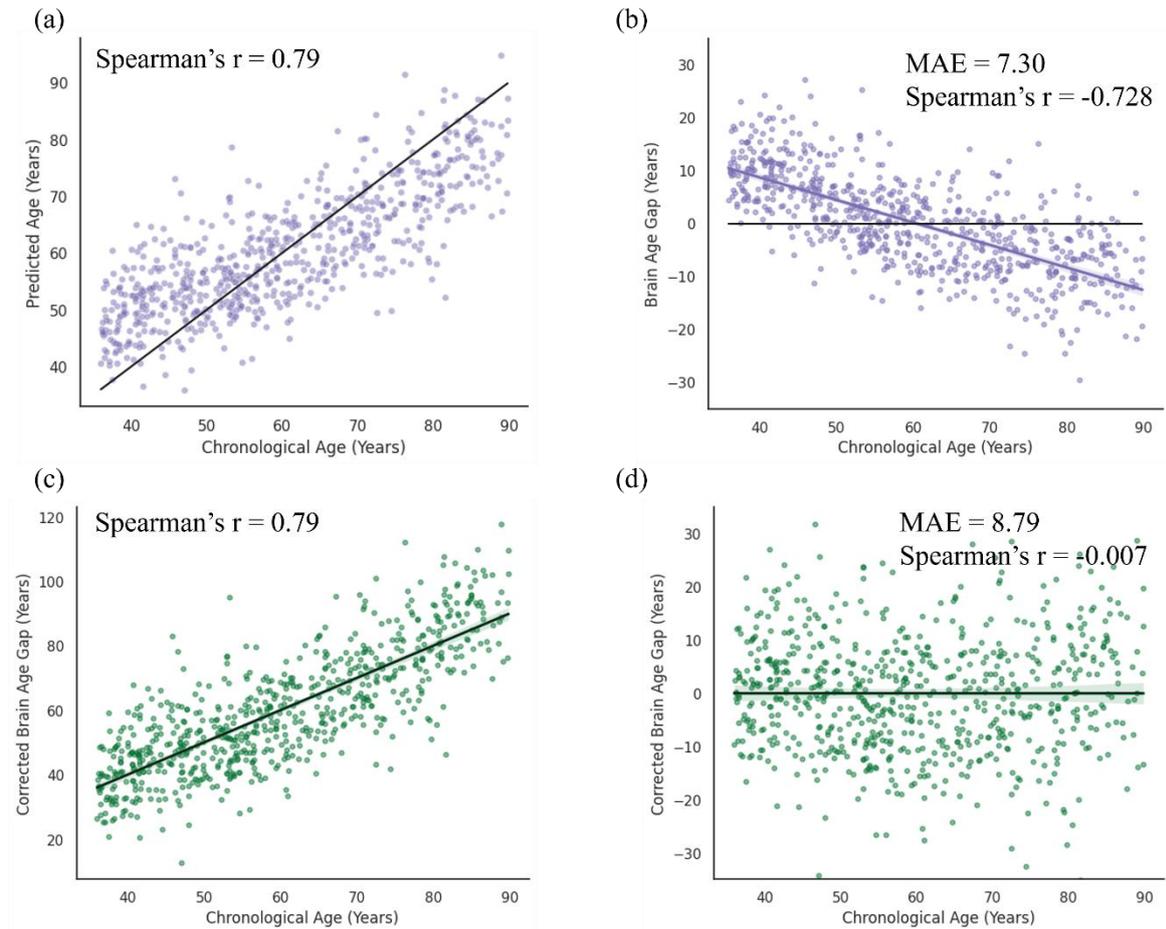

**Suppl Fig. 3.** Age prediction and brain age gap (i.e. predicted age – chronological age) before and after linear correction. (a) shows the predicted age versus chronological age without correction. (b) shows the brain age gap versus age before correction. (c) After correction, the predicted age and chronological age have a similar spearman's correlation. (d) The linear correction removes the negative correlation between the brain age gap and the chronological age. All the black lines in these subfigures show the prefect prediction and the coloured one (either purple or green lines) are the linear fit line of the data.

## 5. Saliency Map Spatial Resolution

The main reason for using layer-wise class activation map (LayerCAM) rather than gradient-weighted class activation mapping (Grad-CAM) is to provide a much better spatial resolution for the saliency maps. Suppl. Fig.4 shows example saliency maps generated using Grad-CAM and LayerCAM on the same subject. The saliency map generated using the LayerCAM method has a much higher spatial resolution, enabling our analysis on a finer detail compared to the saliency map generated using Grad-CAM.

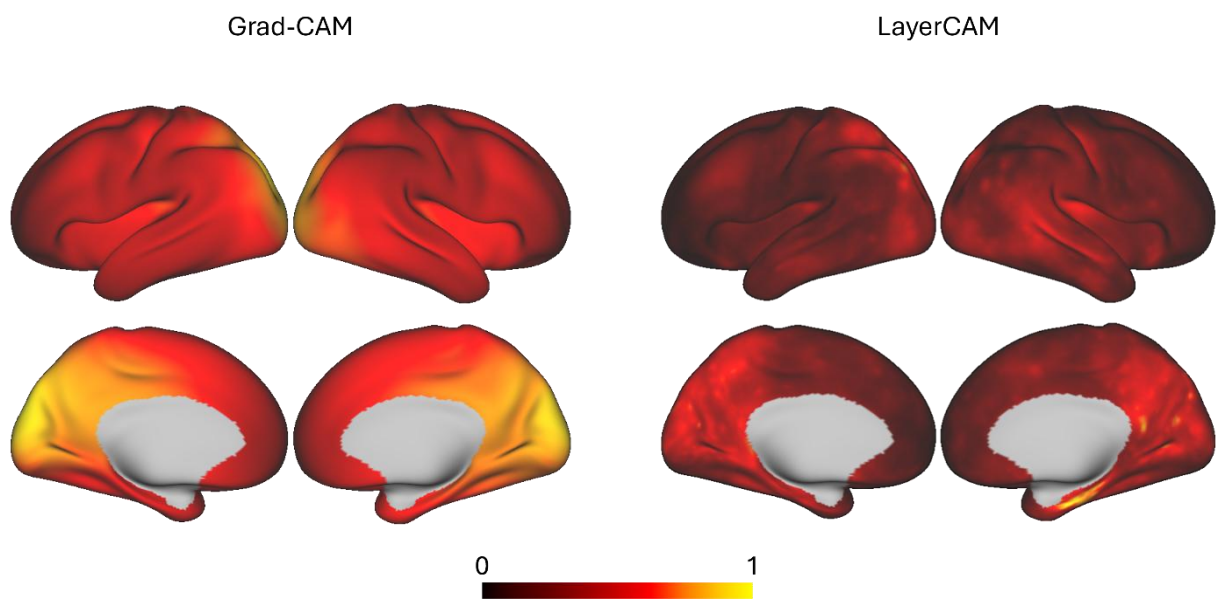

**Suppl. Fig. 4.** Comparison of Grad-CAM and LayerCAM generated saliency maps. This figure shows the two saliency maps generated using Grad-CAM and LayerCAM on the same subject. The saliency maps are mapped to the inflated brain surface. The LayerCAM saliency map shows much higher spatial specificity.

## 6. Regression Analysis on Regions with High Saliency

We investigated the linear relationships in the six regions identified in our saliency maps, including precuneus, cuneus, posterior cingulate cortex (PCC), parahippocampal cortex (PHC), left superior temporal sulcus (right STS), and right superior parietal lobule (right SPL). For each highlighted region, we obtained the averaged hippocampal FC for each subject using the masks shown in Suppl. Fig.4. Based on the regional FC with hippocampus, we applied linear regression, as well as second and third order polynomial regression. We recorded the p-values and $R^2$ as shown in Suppl. Table1. The PHC, left SPL, and cuneus shows relatively stronger linear relationships among these regions, which is aligned with results displayed in Fig2c&f. The linear fit for these regions achieved $R^2$ higher than 0.02. The low explained variance aligns with previous work studying age effects on hippocampal FC [8], where the low adjusted $R^2$ was reported. However, for the other three regions, i.e. right STS, precuneus, and PCC, their $R^2$ are quite low, especially for PCC with only 0.0078, indicating the linear fit is limited to explain the high individual variance well with the aging effect. According to Suppl. Table1, increasing the order of the polynomial regression, the $R^2$ increases in most of the regions, especially for the three regions showing low $R^2$. This implies that the hippocampal FC in these regions can be better explained using more complex and higher order relationships compared to simple linear relationships.

These regional linear regression results show our saliency maps capture regions with relatively significant linear relationships and also ones with weak relationships. This indicates our deep learning methods can differentiate aging-relevant voxel-level FC patterns, acting as a complementary method to identify complex patterns beyond weak linear relationship identified from traditional approaches.

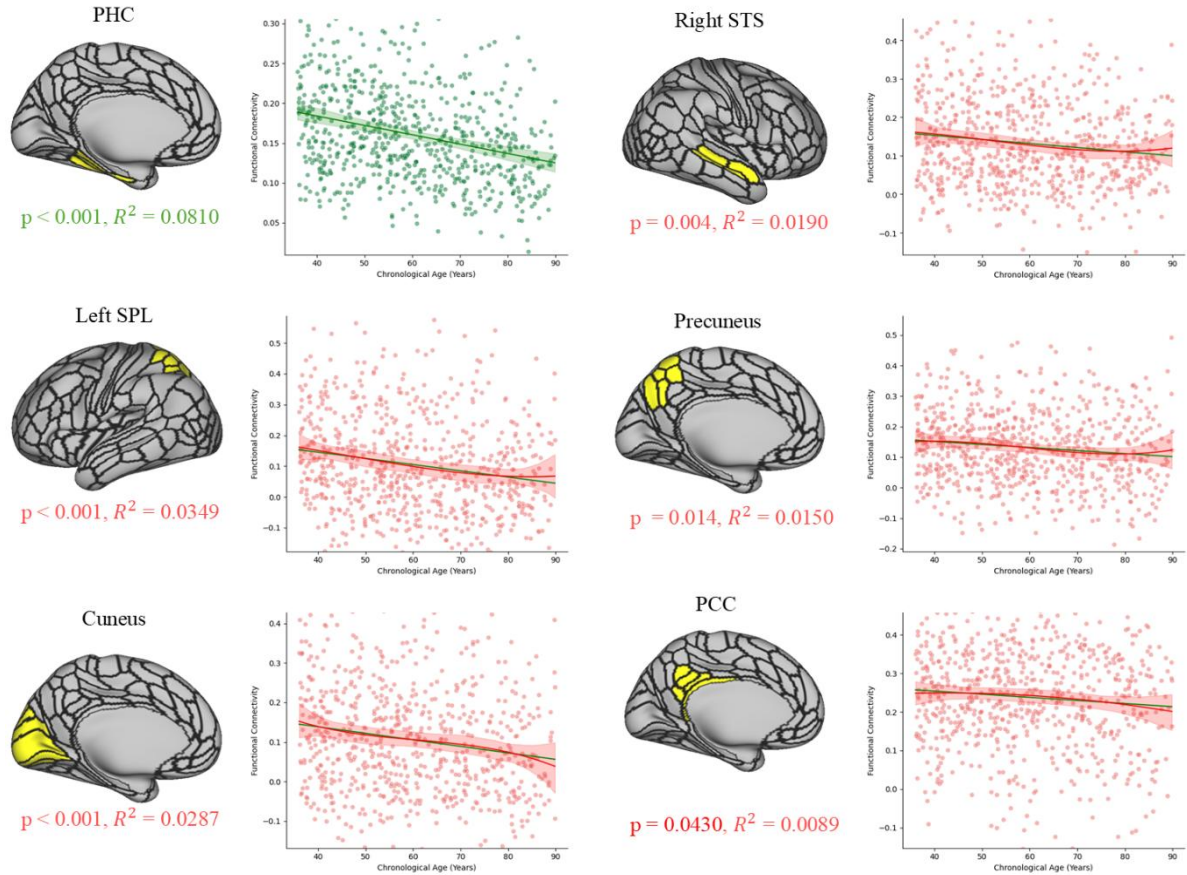

**Suppl. Fig. 5.** Regional based regression results. Yellow regions highlighted on the brain surface with the MMP atlas overlaid shows the corresponding mask used to calculate the averaged FC. Scatter plots on the right-hand side for each regional mask show the corresponding functional connectivity (y-axis) versus chronological age (x-axis). The left two columns show three regions with relatively high $R^2$ in their regression relationship. On the right-hand side, we show the other three regions with relatively low $R^2$. The results presented in green are based on linear regression, while the red ones are high-order polynomial regression results. Green lines in the red scatter plots are still the linear fit lines.

**Suppl. Table 2** Significant models and their coefficient of determination ($R^2$) from linear and higher-order polynomial regression for the six regions highlighted in the averaged saliency map. P-values shows the significance of the overall model. $R^2$ explains the proportion of the explained variance. Higher values mean the model can explain more variance in the dependent variable (functional connectivity).

| Region | Linear Regression | | Second-Order Polynomial Regression | | Third-Order Polynomial Regression | |
| --- | --- | --- | --- | --- | --- | --- |
| | p | $R^2$ | p | $R^2$ | p | $R^2$ |
| PHC | <0.001 | 0.0810 | <0.001 | 0.0810 | <0.001 | 0.0810 |
| Left SPL | <0.001 | 0.0336 | <0.001 | 0.0348 ↑ | <0.001 | 0.0349 ↑ |
| Cuneus | <0.001 | 0.0279 | <0.001 | 0.0280 | <0.001 | 0.0287 ↑ |
| Right STS | <0.001 | 0.0175 | 0.0013 | 0.0187 ↑ | 0.0036 | 0.0190 ↑ |
| Precuneus | 0.0018 | 0.0137 | 0.0069 | 0.0140 ↑ | 0.0136 | 0.0150 ↑ |
| PCC | 0.0184 | 0.0078 | 0.0430 | 0.0089 ↑ | 0.0580 | 0.0106 |